\documentclass[lettersize,journal]{IEEEtran}

\usepackage[colorlinks,urlcolor=blue,linkcolor=blue,citecolor=blue]{hyperref}
  \usepackage{cite}
\usepackage{csquotes}
\usepackage{algorithm}
\usepackage{algpseudocode}
\usepackage{epsfig}
\usepackage{graphicx}
\usepackage{amsfonts}
\usepackage{comment}
\usepackage{color}
\usepackage{amssymb}
\usepackage{amsmath, mathtools}
\usepackage{amsthm}
\usepackage{etoolbox}
\usepackage{flushend}
\usepackage{babel,blindtext}
\usepackage{caption}
\usepackage{subcaption}
\usepackage{csquotes}
\usepackage{rotating}
\usepackage{textcomp}
\usepackage{footnote}
\usepackage{tablefootnote}
\usepackage[flushleft]{threeparttable}
\usepackage{paralist}
\usepackage{mdwlist}
\usepackage{url}
\usepackage{verbatim}
\usepackage{alltt}
\usepackage{multirow}
\usepackage{epstopdf}
\usepackage{lipsum}
\usepackage{float}
\usepackage{booktabs}
\usepackage{slashbox}
\usepackage[table]{xcolor}
\usepackage{array}
\usepackage{boldline}
\usepackage{caption,setspace}
\usepackage{stackengine}[2013-10-15]

\usepackage[T1]{fontenc}
\usepackage{frcursive}
\usepackage{calligra}
\usepackage{wedn}
\usepackage{bm}
\usepackage{aurical}
\usepackage{upgreek}

\definecolor{darkgreen}{RGB}{0,204,0}

\usepackage{cite}
\usepackage{adjustbox}
\usepackage{esint}
\usepackage{tipa}

\usepackage{tikz}
\usepackage{pgfmath}
\usepackage{pgfplots}
\pgfplotsset{compat=1.18}
\usepackage[dvipsnames]{xcolor}
\usetikzlibrary{calc}
\usetikzlibrary{matrix} 

\definecolor{darkgreen}{RGB}{0,153,0}
\definecolor{darkred}{RGB}{192,0,0}

\usepackage[T1]{fontenc}

\newcolumntype{?}{!{\vrule width 1pt}}
\newcolumntype{+}{!{\vrule width 1.25pt}}

\def\hlineb#1{%
\noalign{\ifnum0=`}\fi\hrule \@height #1 %
\futurelet\reserved@a\@xhline}

\usepackage{cancel}

\definecolor{lightgray}{gray}{0.8}

\usepackage{stackengine}[2013-10-15]
\usepackage{ dsfont }
\usepackage[T1]{fontenc}
\usepackage{frcursive}
\usepackage{calligra}
\usepackage{wedn}
\usepackage{bm}
\usepackage{aurical}
\usepackage{ upgreek }

\usepackage{adjustbox}
\usepackage{xcolor}
\usepackage{makecell}
\usepackage{ mathrsfs }

\usepackage{pgfplots}
\pgfplotsset{compat=1.18}

\usepackage{csquotes}

\ifCLASSINFOpdf
\else
\fi
\begin{document}


\title{Adaptive Data-Resilient Multi-Modal Hierarchical Multi-Label Book Genre Identification}

\author{Utsav Kumar Nareti, 
Soumi Chattopadhyay, 
Prolay Mallick, 
Suraj Kumar, 
Chandranath Adak, 
Ayush Vikas Daga, 
Adarsh Wase, 
Arjab Roy 
\thanks{
U.K. Nareti and C. Adak are with the Dept. of CSE, IIT Patna, India. 
S. Chattopadhyay, P. Mallick, and S. Kumar are with the Dept. of CSE, IIT Indore, India. 
A. V. Daga is with SVNIT, India. 
A. Wase is with IIT Indore and IIM Indore, India. 
A. Roy is with the Dept. of HSS, IIIT Guwahati, India. 
%
%
%
%
%
\emph{Corresponding authors:} C. Adak, S. Chattopadhyay.
}
}



\markboth{U. K. Nareti \MakeLowercase{\textit{et al.}}}
{U. K. Nareti \MakeLowercase{\textit{et al.}}: XXX}

\maketitle

\begin{abstract}
Identifying fine-grained book genres is essential for enhancing user experience through efficient discovery, personalized recommendations, and improved reader engagement. At the same time, it provides publishers and marketers with valuable insights into consumer preferences and emerging market trends. While traditional genre classification methods predominantly rely on textual reviews or content analysis, the integration of additional modalities, such as book covers, blurbs, and metadata, offers richer contextual cues. However, the effectiveness of such multi-modal systems is often hindered by incomplete, noisy, or missing data across modalities.
To address this, we propose IMAGINE (\emph{Intelligent Multi-modal Adaptive Genre Identification NEtwork}), a framework designed to leverage multi-modal data while remaining robust to missing or unreliable information. IMAGINE learns modality-specific feature representations and adaptively prioritizes the most informative sources available at inference time. It further employs a hierarchical classification strategy, grounded in a curated taxonomy of book genres, to capture inter-genre relationships and support multi-label assignments reflective of real-world literary diversity.
A key strength of IMAGINE is its adaptability: it maintains high predictive performance even when one modality, such as text or image, is unavailable. We also curated a large-scale hierarchical dataset that structures book genres into multiple levels of granularity, allowing for a more comprehensive evaluation. Experimental results demonstrate that IMAGINE outperformed strong baselines in various settings, with significant gains in scenarios involving incomplete modality-specific data.

\end{abstract}

\begin{IEEEkeywords}
Hierarchical classification, Multi-label classification, Multi-modal classification, Adaptive learning 
\end{IEEEkeywords}

\section{Introduction}\label{sec:intro} 

\IEEEPARstart{I}{n} the digital media landscape, accurate book genre classification is central to effective recommendations, enhancing user experience and engagement on literary platforms. It enables readers to discover books aligned with their preferences and provides publishers and marketers with insights into consumer behavior, content curation, and targeted marketing. By refining categorization, genre-based recommendations enrich user interactions and support informed decisions in book production, promotion, and distribution \cite{DBLP:conf/wise/NgJ19}. 
The rise of eBooks and digital reading has further transformed publishing, enabling global distribution via platforms like Goodreads and Amazon Kindle \cite{lee2023can}. While this expansion increases accessibility, it also creates challenges in navigating vast digital libraries, making manual classification impractical and underscoring the need for automated genre identification. Goodreads addresses this through user-generated 
shelves, 
but this approach depends on unreliable reviews, fails when reviews are absent, and often produces non-standard (e.g., book format such as {audiobook} or {paperbook}) or conflicting labels (e.g., \textit{fiction} and \textit{non-fiction}). Moreover, it lacks a hierarchical structure, limiting organization across broad and fine-grained genres. 

\begin{figure}
\centering
\includegraphics[width=\linewidth]{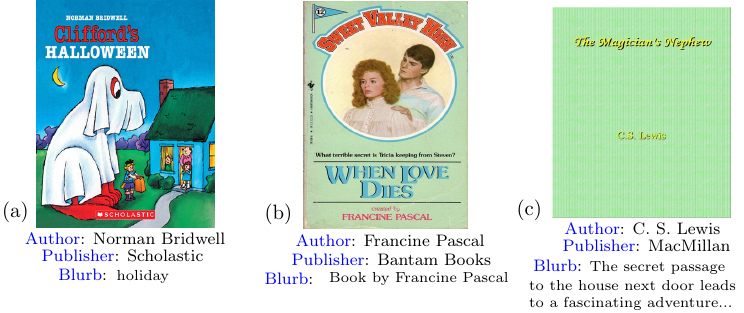}
\caption{Examples of content discrepancies: (a) Uninformative blurb, (b) Irrelevant blurb, (c) Minimal visual cues on cover}
\label{fig:example_intro}
\end{figure}

Despite its importance, automated book genre classification remains underexplored. Prior studies rely mainly on single modalities such as descriptions \cite{DBLP:conf/icmmi/SobkowiczKB17,DBLP:conf/hicss/Khalifa022} or cover images \cite{article_2,DBLP:conf/icpram/BuczkowskiSK18}. Some multi-modal methods combine metadata with cover images \cite{DBLP:journals/ijdar/RasheedUSKS23,article}, but they typically achieve low accuracy and ignore hierarchical structures crucial for nuanced genre relationships. As shown in Fig.~\ref{fig:example_intro}, books may have minimal descriptions, sparse metadata, or uninformative covers. These limitations highlight the need for a comprehensive multi-modal framework that integrates cover images, blurbs, and metadata for structured genre identification, thereby strengthening recommendation systems and user experience.

A further drawback of prior works \cite{saraswat2022leveraging} is their reliance on inconsistent user-generated reviews or labels. To overcome this, we construct a multi-modal dataset that aggregates reliable sources: cover images, blurbs, metadata, and OCR-extracted cover text. Unlike datasets based solely on crowd-sourced labels, ours employs expert-verified annotations organized into a hierarchical taxonomy, enabling precise classification. To mitigate class imbalance, we adopt a two-stage preprocessing strategy with data augmentation and selective resampling.

Building on this foundation, we present IMAGINE, an adaptive multi-modal framework for hierarchical multi-label genre classification. 
Our key \textbf{contributions} are:

\emph{\textbf{(i)} Hierarchical multi-modal formulation of book genre identification}: Unlike movie genres, book genre classification is comparatively underexplored. Prior works rely on single modalities (e.g., cover images, titles, blurbs, or reviews) or limited pairs (cover image + title). To the best of our knowledge, we are the first to formulate hierarchical book genre classification, where Level-1 distinguishes {fiction} vs. {non-fiction}, and Level-2 performs fine-grained multi-label classification. 

\emph{\textbf{(ii)} IMAGINE: An adaptive, data-resilient multi-modal framework}: We propose IMAGINE, a novel framework with four methodological innovations: (a) a two-level hierarchy for broad-to-fine multi-label genre identification, (b) comprehensive multi-modal fusion that captures richer context across diverse inputs, (c) a selective gating mechanism that adaptively prioritizes the most informative modality under noisy or missing data, and (d) an imbalance-aware loss function that mitigates skewed label distributions, improving robustness for underrepresented genres.

\emph{\textbf{(iii)} New dataset and benchmarking}: 
We construct a dataset of 11302 book samples comprising cover images, blurbs, metadata with expert-verified hierarchical multi-label genres annotations, on which we conduct extensive experiments benchmarking IMAGINE against state-of-the-art unimodal, multi-modal, hierarchical, and large-scale models. Detailed analyses, including ablation studies and genre-wise evaluations, demonstrate that IMAGINE consistently outperforms baselines in both accuracy and adaptability, establishing a new benchmark for structured and reliable book genre classification.



The paper is organized as follows: Section \ref{sec:related_work} reviews related work, Sections \ref{sec:prob_form}-\ref{sol_arch} present the problem and IMAGINE’s architecture, Section \ref{4sec:result} reports experiments,  Section \ref{5sec:conclusion} concludes, and the supplementary file provides dataset details, challenges, augmentation strategies, and qualitative results.

\section{Related Work}\label{sec:related_work}

This paper primarily focuses on identifying multi-label book genres using multi-modal data. 
Prior studies explored various modalities in isolation or
combination, which we outline in Table \ref{tab:related_work}, and briefly summarize below.

\textbf{\em Visual:} Cover images serve as the visual representation of a book, incorporating elements such as visual scenes, titles, font styles, and illustrations, all of which provide meaningful cues about the book's genre.
CNNs were utilized in \cite{iwana2016judging} to analyze book cover images for classifying into 30 genres.
Similarly, in \cite{article_2}, various CNN-based architectures were engaged to  classify into 32 genres based on cover images.
In \cite{DBLP:conf/icpram/BuczkowskiSK18}, CNNs were also used to classify books into 14 genres.

\textbf{\em Textual:} Textual data in books, including blurbs, titles, metadata, and user-generated reviews, provides rich information for genre identification.
In  \cite{DBLP:conf/icmmi/SobkowiczKB17}, Naïve Bayes and Doc2Vec were used to analyze blurbs for genre classification.
In \cite{saraswat2022leveraging}, an RNN with LSTM was employed to categorize book reviews into genres, whereas RNN with GRU was engaged in \cite{DBLP:conf/wise/NgJ19} to analyze blurbs. 
CNNs  \cite{DBLP:conf/hicss/Khalifa022} incorporating pre-trained universal sentence encoder (USE) processed book content for predicting genres. 
In \cite{spanish_book}, BERT was applied to classify the blurb of Spanish books.
CNN-LSTM with attention was employed in \cite{bangla_book} on the Bangla book blurb.
Portuguese user reviews with TF-IDF/ LSA features were engaged in \cite{port_book} for genre identification.

\textbf{\em Multi-modal:}
Multiple modalities, such as cover images, blurbs, reviews, and metadata, have often been combined to enhance book genre prediction accuracy.
A multi-modal model integrating ResNet-50 for cover image and USE for cover text was proposed in  \cite{DBLP:journals/corr/abs-2011-07658} to classify genres. Similarly, \cite{Biradar} utilized XceptionNet for extracting features from cover images and GloVe embeddings from titles to feed into a multinomial logistic regression model.
A multi-modal attention fusion framework, employing a modified SE-ResNeXt handled cover image, book title, and metadata \cite{DBLP:journals/ijdar/RasheedUSKS23}. Inception-v3 and Na\"{i}ve Bayes were used in \cite{article} to extract features from the cover image and cover text, respectively, and then fused using early and late fusion techniques to enhance genre classification.

\begin{table}[!t]
 \centering
 \caption{\small{Summary of related work on book genre identification}}
   \begin{adjustbox}{width=0.47\textwidth} 
 \begin{tabular}{c | c | c | c | c | c | c}
   \cline{2-7} 
\multicolumn{1}{c}{} & \multirow{2}{*}{{\textbf{Method}}} &  \multirow{2}{*}{\textbf{Input}} & \textbf{Architecture}/  & \#\textbf{Genre}/ & \multirow{2}{*}{\textbf{Dataset}} & {{\textbf{Multi}-}}\\
\multicolumn{1}{c}{} & & & \textbf{Technique} & \#\textbf{Tags} & & {{\textbf{label?}}} \\  


\hline \hline 
\multirow{3}{*}{\rotatebox{90}{Visual}}  & \cite{iwana2016judging} & Cover image & AlexNet & 30 & Amazon & $\upchi$\\ \cline{2-7}
 & \cite{article_2} & Cover image & CNN & 32 & Amazon  & $\upchi$\\ \cline{2-7}
  & \cite{DBLP:conf/icpram/BuczkowskiSK18} & Cover image & CNN & 14 & GoodReads  ($\wp$) & $\upchi$\\  


\hline \hlineB{2} 
\multirow{8}{*}{\rotatebox{90}{Textual}} &  \cite{DBLP:conf/icmmi/SobkowiczKB17}  & Blurb & Naive Bayes, Doc2Vec & 14 & GoodReads  ($\wp$)  & $\upchi$ \\ \cline{2-7}

&  \multirow{1}{*}{\cite{DBLP:conf/wise/NgJ19}}  & Blurb, Reviews, Rating & \multirow{1}{*}{RNN + GRU} & \multirow{1}{*}{31} & \multirow{1}{*}{Book-Crossing  ($\wp$)} & \multirow{1}{*}{$\upchi$} \\ 
  \cline{2-7}

& \multirow{2}{*}{\cite{saraswat2022leveraging}}   & \multirow{2}{*}{Reviews} & \multirow{2}{*}{RNN + LSTM} & \multirow{2}{*}{28} & Book-Crossing  ($\wp$), & \multirow{2}{*}{\checkmark} \\ 
& &  & & & Amazon & \\ \cline{2-7}
&  \multirow{1}{*}{\cite{DBLP:conf/hicss/Khalifa022}}  & \multirow{1}{*}{Book content}  & USE, CNN & \multirow{1}{*}{8} & \multirow{1}{*}{GoodReads  ($\wp$)}  & \multirow{1}{*}{$\upchi$} \\ \cline{2-7}

& \cite{bangla_book} & Blurb & CNN, LSTM, Attention & 8 & Bangla Book Dataset ($\wp$) & \checkmark \\ \cline{2-7}
& \cite{spanish_book} & Blurb & BERT & 26 & Spanish Book Dataset ($\wp$) & $\upchi$ \\ \cline{2-7}
& \cite{port_book} & Reviews & TF-IDF, Random Forest & 24 & Portuguese Books ($\wp$)  & $\upchi$ \\ \hline


\hline \hlineB{2} 
\multirow{6}{*}{\rotatebox{90}{Multi-modal}}
&  \multirow{1}{*}{ \cite{chiang2015classification}}  & Cover image, Book title & \multirow{1}{*}{CNN, NLP, SVM} & \multirow{1}{*}{5} & \multirow{1}{*}{OpenLibrary.org  ($\wp$)}  & \multirow{1}{*}{$\upchi$} \\  
  \cline{2-7}

& \multirow{1}{*}{\cite{DBLP:journals/corr/abs-2011-07658}}  & \multirow{1}{*}{Cover image, Cover text}  & USE, ResNet50 & \multirow{1}{*}{30} & \multirow{1}{*}{BookCover30  ($\wp$)} & 
   \multirow{1}{*}{$\upchi$} \\ \cline{2-7}

& \multirow{1}{*}{\cite{Biradar}}  & Cover image, Book title  & \multirow{1}{*}{Xception, GloVe} & \multirow{1}{*}{5} & \multirow{1}{*}{Amazon  ($\wp$)} & 
   \multirow{1}{*}{$\upchi$} \\   
    \cline{2-7}

& \multirow{2}{*}{\cite{DBLP:journals/ijdar/RasheedUSKS23}}  & Cover image, Book title,  & \multirow{2}{*}{SE-ResNeXt-101, EXAN} & \multirow{2}{*}{28} & BookCover28, & 
   \multirow{2}{*}{$\upchi$} \\   
    & &Metadata & & & Arabic Book Cover  ($\wp$) & \\ \cline{2-7}

& \multirow{1}{*}{\cite{article}}  & Cover image, Book title  & Inception-v3, Naive Bayes, & \multirow{1}{*}{30} & \multirow{1}{*}{Amazon} & 
   \multirow{1}{*}{$\upchi$} \\   
    \cline{2-7}

    \hline


  %
  \multicolumn{7}{r}{($\wp$): Publicly unavailable}
 \end{tabular}
 \end{adjustbox}
 
 \label{tab:related_work}
\end{table}

\textbf{\emph{Positioning of Our Work:}} 
Existing literature on book genre identification remains limited, with most studies focusing on either visual or textual inputs.
A few works have explored multi-modal data combining cover images and text, yet none have systematically leveraged all key book modalities, cover page, cover text, metadata, and blurb, within a unified framework. Moreover, prior efforts rarely address the challenges of multi-label classification or the hierarchical nature of book genres.

In contrast, IMAGINE is 
the earliest of its kind
to perform hierarchical multi-label book genre classification using comprehensive multi-modal data. It introduces a selective gating module that dynamically selects the most informative modality, making it robust to missing or incomplete inputs. Additionally, IMAGINE offers genre-wise performance insights and an in-depth misprediction analysis, aspects largely overlooked in existing research. These contributions position IMAGINE as a significant advancement in multi-modal, hierarchical genre classification.

\section{Problem Formulation}
\label{sec:prob_form}

We define the hierarchical book genre classification problem as follows.  
Let ${\cal B}=\{{\cal B}_1,\ldots,{\cal B}_n\}$ be a collection of $n$ books, where each book is represented as:  
\(
{\cal B}_i = ({\cal I}_i, {\cal T}_i, {\cal M}_i),
\)
with ${\cal I}_i$ denoting the cover image, ${\cal T}_i$ the blurb, and ${\cal M}_i$ the structured metadata. 
Genres are organized hierarchically:
\[\scriptsize
L_1=\{0,1\}, ~~ 
L_f=\{\gamma_1,\ldots,\gamma_{m_1}\}, ~~ 
L_{nf}=\{\lambda_1,\ldots,\lambda_{m_2}\},
\]
where, `0' corresponds to \emph{fiction} and `1' to \emph{non-fiction}. 
$\gamma_i \in L_f$ refers to a genre corresponding to fiction, and $\lambda_i \in L_{nf}$ refers to a genre corresponding to non-fiction. 
Each book ${\cal B}_i$ is associated with a label:
\[\scriptsize
{\cal Y}_i = ({\cal Y}^i_1, {\cal Y}^i_2), \quad {\cal Y}^i_1 \in L_1, \quad
{\cal Y}^i_2 \subseteq
\begin{cases}
L_f, & \text{if } {\cal Y}^i_1=0 ~ (\text{fiction}), \\
L_{nf}, & \text{if } {\cal Y}^i_1=1 ~ (\text{non-fiction}).
\end{cases}
\]
Thus, {Level-1} is a binary classification task (\emph{fiction vs. non-fiction}), while {Level-2} is a conditional multi-label classification task within the selected branch. The goal is to learn a function:
\[\scriptsize
f: ({\cal I}_i, {\cal T}_i, {\cal M}_i) \mapsto ({\cal Y}^i_1, {\cal Y}^i_2),
\]
that predicts the hierarchical genres of unseen books by leveraging multi-modal inputs, while remaining robust to noisy or missing modalities.

\section{Solution Architecture}
\label{sol_arch}

\begin{figure*}
    \centering
    \includegraphics[width=\linewidth]{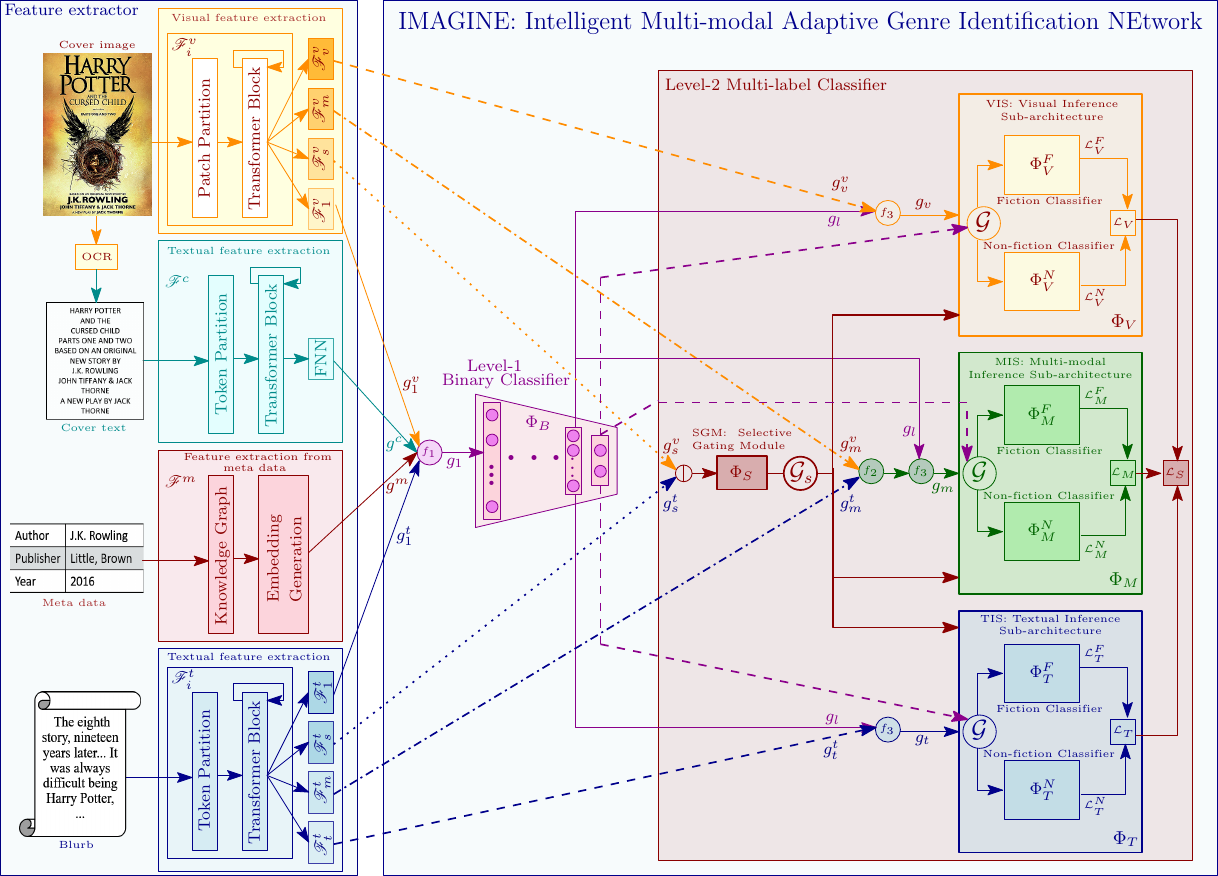}
    \caption{Overview of our framework IMAGINE for hierarchical book genre prediction}
    \label{fig:wf}
\end{figure*}

This section presents our proposed framework, IMAGINE, for hierarchical multi-modal book genre prediction (Fig.~\ref{fig:wf}). IMAGINE leverages four input sources: cover image, OCR-extracted cover text, blurb, and metadata (e.g., author, publisher), which are processed through three alternative processing pathways: a visual pathway ($\psi_V$), a textual pathway ($\psi_T$), and a multi-modal fusion pathway ($\psi_M$).

At the core of IMAGINE lies the selective gating module ($\Phi_S$), which evaluates the availability and reliability of inputs, and then activates the most suitable pathway. Unlike conventional multi-modal fusion approaches that always combine all modalities, $\Phi_S$ selects only one pathway at a time, ensuring both efficiency and robustness when some modalities are missing or noisy. Regardless of the chosen pathway, all follow a two-stage hierarchical classification process: 
(\emph{a}) Level-1: A shared classifier ($\Phi_B$) aggregates all features to determine whether a book is \emph{fiction} or \emph{non-fiction}, 
(\emph{b}) Level-2: Conditioned on $\Phi_B$’s decision, a pathway-specific module refines the classification into fine-grained genres. 

Specifically, on the visual pathway ($\psi_V$), $\Phi_B$ is followed by $\Phi_V$, which integrates the features of the cover image with the latent representation $g_l$ of $\Phi_B$. In the textual pathway ($\psi_T$), $\Phi_B$ is followed by $\Phi_T$, which processes the blurb together with 
$g_l$. In the multi-modal pathway ($\psi_M$), $\Phi_B$ is followed by $\Phi_M$, which fuses cover image and blurb features with 
$g_l$.

Each Level-2 module ($\Phi_i$, $i \in \{V,T,M\}$) includes two parallel classifiers: $\Phi_i^F$ for \emph{fiction} sub-genres and $\Phi_i^N$ for \emph{non-fiction} sub-genres. The Level-1 prediction from $\Phi_B$ activates the appropriate classifier, ensuring category-aware and fine-grained genre identification. 
The subsequent subsections describe the architecture of each module in detail.

\subsection{SGM: Selective Gating Module ($\Phi_S$)}
The top-level module of IMAGINE, denoted as $\Phi_S$, functions as a {selective gating mechanism} that dynamically routes each book to the most suitable pathway, visual ($\psi_V$), textual ($\psi_T$), or multi-modal ($\psi_M$), based on the reliability and completeness of its inputs. This design enables IMAGINE to remain robust under noisy, incomplete, or modality-specific conditions.  
$\Phi_S$ operates on concatenated feature embeddings: visual features ($g_s^v$) from the cover image and textual features ($g_s^t$) from the blurb. Implemented as a deep feedforward neural network, it produces a probability distribution $\hat{Y} \in \mathbb{R}^3$ over the three pathways. A gating function $\mathcal{G}_s$ converts this distribution into a one-hot routing decision by selecting the pathway with maximum confidence:  
\begin{equation}
\mathcal{G}_s(\hat{Y}) = \mathbf{e}_{\{\arg\max_j \hat{Y}_j\}}, \quad j \in \{V,T,M\}
\end{equation}
where, $\mathbf{e}_i$ denotes the one-hot vector corresponding to the chosen pathway. Thus, only one pathway is activated during both training and inference, promoting specialization across modality-specific branches.  

Training of $\Phi_S$ is guided by a cross-entropy loss ($\mathcal{L}_S$), encouraging robust and context-aware routing. Furthermore, $\Phi_S$ employs {experience-based supervision}, where each training instance is assigned to the pathway that previously yielded the most accurate prediction. This feedback-driven adaptation ensures that the gating mechanism evolves in alignment with empirical performance, leading to more reliable and accurate multi-label genre classification.  

\subsection{MIS: Multi-modal Inference Sub-Architecture ($\psi_M$)} 
MIS employs a two-level hierarchical structure: a Level-1 binary classifier ($\Phi_B$) and a Level-2 multi-label classifier ($\Phi_M$). These levels are connected through gating, where $\Phi_B$ determines the broad category (fiction vs. non-fiction) and activates the corresponding Level-2 branch for fine-grained classification.

\subsubsection{Level-1 Binary Classifier ($\Phi_B$)}  
The first stage of IMAGINE is a shared binary classifier $\Phi_B$, applied uniformly across all pathways ($\psi_V$, $\psi_T$, $\psi_M$). Its task is to distinguish between \emph{fiction} and \emph{non-fiction}. To this end, $\Phi_B$ aggregates features from multiple modalities: visual features ($g_1^v$) extracted from the cover image, textual features from the blurb ($g_1^t$) and cover text ($g^c$), and metadata features ($g^m$) such as author or publisher information. These are fused into a joint representation:
\(
g_1 = f_1(g_1^v, g_1^t, g^c, g^m),
\),
where $f_1$ denotes the fusion function. Empirically, simple concatenation consistently outperformed more complex schemes (e.g., linear projections, self- and cross-attention), and is therefore adopted throughout.  

The aggregated feature $g_1$ is fed into $\Phi_B$, a feedforward neural network optimized with binary cross-entropy (BCE), producing a probability vector $\hat{\mathcal{Y}}_1$. A gating function $\mathcal{G}$ then converts $\hat{\mathcal{Y}}_1$ into a one-hot routing decision, selecting either the fiction or non-fiction branch for Level-2 classification. Unlike the top-level selective gating module $\Phi_S$, which chooses the most informative modality pathway, $\mathcal{G}$ governs branch activation 
within
Level-2 according to $\Phi_B$’s prediction.  

\subsubsection{Level-2 Multi-label Classifier ($\Phi_M$)}  
Conditioned on the Level-1 output, the second stage $\Phi_M$ refines predictions into fine-grained genres through multi-label classification. It consists of two parallel classifiers: $\Phi^F_M$ for \emph{fiction} sub-genres and $\Phi^N_M$ for \emph{non-fiction} sub-genres. The active branch is determined by $\mathcal{G}(\hat{\mathcal{Y}}_1)$.  
Each classifier integrates multi-modal information by combining visual cover features ($g_m^v$), textual blurb features ($g_m^t$), and the latent representation $g_l$ from $\Phi_B$. These are fused as:  
\(
g_m = f_3(f_2(g_m^v, g_m^t), g_l)
\),
where $f_2$ merges modality-specific features and $f_3$ incorporates Level-1 context. 
Here also, concatenation proved most effective.

Both $\Phi^F_M$ and $\Phi^N_M$ are implemented as feedforward neural networks. A sigmoid activation is applied at the final layer to generate multi-label probability estimates $\hat{\mathcal{Y}}_2$. A genre is assigned when its probability exceeds an empirical threshold.

Training uses the asymmetric loss function (ASL) \cite{asl_loss}, designed for multi-label imbalance. For each sample $i$, the \emph{fiction} and \emph{non-fiction} losses are:  
\begin{equation} \label{eq:mm_loss}
\scriptsize
\begin{aligned}
{\mathcal{L}}_M^{F(i)} = \frac{1}{m_1} \sum_{j=1}^{m_1} 
\big( {\mathcal{L}}_{MF}^{ij+} + {\mathcal{L}}_{MF}^{ij-} \big);~ 
{\mathcal{L}}_M^{N(i)} = \frac{1}{m_2} \sum_{j=1}^{m_2} 
\big( {\mathcal{L}}_{MN}^{ij+} + {\mathcal{L}}_{MN}^{ij-} \big)
\end{aligned}
\end{equation} 
Here, $m_1$ and $m_2$ denote the number of fiction and non-fiction genres, respectively.
The positive term emphasizes underconfident true labels:  ${\mathcal{L}}_{MF}^{ij+} = {\mathcal{Y}}^{ij}_2 (1-\hat{\mathcal{Y}}^{ij}_2)^{\gamma^+} 
\log(\hat{\mathcal{Y}}^{ij}_2 + \epsilon_0)$,
while the negative term penalizes overconfident irrelevant labels with clipping: ${\mathcal{L}}_{MF}^{ij-} = (1-{\mathcal{Y}}^{ij}_2) P_\epsilon(\hat{\mathcal{Y}}^{ij}_2)^{\gamma^-} 
\log(1-P_\epsilon(\hat{\mathcal{Y}}^{ij}_2))$. 
Here, ${\mathcal{Y}}^{ij}_2$ and $\hat{\mathcal{Y}}^{ij}_2$ are the ground-truth and predicted values, respectively, and 
$P_\epsilon(\hat{\mathcal{Y}}^{ij}_2) = \max(\hat{\mathcal{Y}}^{ij}_2-\epsilon, 0)$.  
The hyperparameters $\gamma^+$, $\gamma^-$, $\epsilon$, and $\epsilon_0$ control the focus on hard positives, suppression of negatives, clipping threshold, and numerical stability, respectively. Similarly, ${\mathcal{L}}_{MN}^{ij+}$ and ${\mathcal{L}}_{MN}^{ij-}$ can be computed.

This hierarchical formulation ensures branch-specific, label-sensitive learning. By combining $\Phi_B$’s category-level routing with ASL-driven multi-label refinement, $\Phi_M$ achieves robust fine-grained genre identification under class imbalance.

\subsection{VIS: Visual Inference Sub-Architecture ($\psi_V$)} 
VIS is another key pathway in IMAGINE, structured hierarchically like MIS. It comprises the shared Level-1 binary classifier ($\Phi_B$) and a Level-2 multi-label visual classifier ($\Phi_V$), linked through a gating mechanism that routes information based on the prediction of $\Phi_B$.
Unlike the multi-modal pathway, which integrates multiple sources, the visual pathway processes only cover image features, combined with the latent representation $g_l$ from $\Phi_B$. Incorporating $g_l$ injects contextual knowledge from the fiction vs. non-fiction decision, improving genre predictions within the visual domain.
The loss functions, $\mathcal{L}_V^{F(i)}$ and $\mathcal{L}_V^{N(i)}$, adopt the same asymmetric formulation as $\mathcal{L}_M^{F(i)}$ and $\mathcal{L}_M^{N(i)}$, but are tailored to the visual pathway.


\subsection{TIS: Textual Inference Sub-Architecture ($\psi_T$)} 
TIS forms the textual pathway of IMAGINE, adopting the same hierarchical structure as MIS and VIS. It consists of the shared Level-1 binary classifier ($\Phi_B$) and a Level-2 multi-label textual classifier ($\Phi_T$), connected through a gating mechanism guided by $\Phi_B$’s output.
As described earlier, $\Phi_B$ is shared across all pathways and determines whether a book belongs to the fiction or non-fiction category. Conditioned on this output, $\Phi_T$ specializes in fine-grained classification within the textual modality. Each classifier within $\Phi_T$ processes blurb features together with the latent representation $g_l$ derived from $\Phi_B$. In this pathway, $g_l$ functions as a conditioning signal from the Level-1 decision, aligning textual representations with the fiction or non-fiction context and thereby enabling more discriminative genre predictions.
Incorporating $g_l$ provides contextual cues from the fiction vs. non-fiction decision, thereby improving fine-grained genre classification in the textual domain.
The pathway-specific losses, $\mathcal{L}_T^{F(i)}$ and $\mathcal{L}_T^{N(i)}$, follow the asymmetric loss formulation of $\mathcal{L}_M^{F(i)}$ and $\mathcal{L}_M^{N(i)}$, but are adapted to textual features.



\subsection{Overall Loss Function}
The IMAGINE loss jointly supervises {Level-1 binary classification} 
(fiction vs. non-fiction) and {Level-2 multi-label genre prediction} 
across modalities. Its design enforces selective gradient propagation, ensuring 
that only the relevant branch and modality are updated for each training instance. 
Formally, let the modality-level gating function be:
\[\scriptsize
{\delta}^{(i)} = \mathcal{G}_s(\hat{Y}^{\,i}) =
[\delta_M^{(i)}, \delta_V^{(i)}, \delta_T^{(i)}]^\top, \quad 
\delta_M^{(i)}+\delta_V^{(i)}+\delta_T^{(i)}=1,
\]
where, exactly one pathway, multi-modal, visual, or textual, is activated per sample. 
The Level-2 losses are then defined as:
\begin{equation}\label{eq:mm_loss_overall}
\scriptsize
\begin{aligned}
\mathcal{L}_F^{2(i)} &=
\delta_M^{(i)}\,\mathcal{L}_M^{F(i)} \;+\;
\delta_V^{(i)}\,\mathcal{L}_V^{F(i)} \;+\;
\delta_T^{(i)}\,\mathcal{L}_T^{F(i)}, \\[4pt]
\mathcal{L}_N^{2(i)} &=
\delta_M^{(i)}\,\mathcal{L}_M^{N(i)} \;+\;
\delta_V^{(i)}\,\mathcal{L}_V^{N(i)} \;+\;
\delta_T^{(i)}\,\mathcal{L}_T^{N(i)} .
\end{aligned}
\end{equation}
The Level-1 supervision is given by the BCE loss:
\begin{equation}
\scriptsize
\mathcal{L}^{1(i)} = \mathrm{BCE}(\mathcal{Y}_1^i,\ \hat{\mathcal{Y}}_1^i).
\end{equation}
The overall training objective aggregates Level-1 and Level-2 supervision:
\begin{equation}\label{eq:overall_final}
\scriptsize
\mathcal{L} = \frac{1}{n}\sum_{i=1}^n
\Big[
\mathcal{L}^{1(i)}
+ \mathcal{Y}_1^i\,\mathcal{G}(\hat{\mathcal{Y}}_1^i)\,\mathcal{L}_F^{2(i)}
+ (1-\mathcal{Y}_1^i)\,(1-\mathcal{G}(\hat{\mathcal{Y}}_1^i))\,\mathcal{L}_N^{2(i)}
\Big],
\end{equation}
where, $\mathcal{G}(\cdot)$ is the class-level gating function (fiction vs. non-fiction) 
governed by $\Phi_B$, and ${\mathcal{Y}}_1^i$ is the Level-1 ground-truth label.
This formulation introduces two complementary selectivity mechanisms:  
(i) {Class-level gating} ($\mathcal{G}$) routes supervision to the correct branch, 
preventing cross-branch interference, and  
(ii) {Modality-level gating} ($\mathcal{G}_s$) activates exactly one modality 
pathway per instance, avoiding noisy updates from weaker signals and encouraging 
specialization.  
During training, the gating module $\Phi_S$ is supervised via experience-based routing 
labels to learn context-aware decisions. At inference, hard one-hot gating is applied.  
Together, the hierarchical routing at class- and modality-level ensures robust, 
context-sensitive predictions aligned with both the {genre taxonomy} and the 
{multi-modal nature} of input.

\subsection{Feature Extractor} 
\label{subsubsec:feature_extraction}
We now discuss the feature extractor modules of IMAGINE, which are responsible for multi-modal feature extraction. These modules are critical for extracting relevant information from different modalities. 

\subsubsection{Visual Feature Extractor from Cover Image}
IMAGINE employs the Swin transformer \cite{SwinT} as the visual backbone for book cover feature extraction, chosen for its efficiency and ability to capture both fine-grained and hierarchical patterns. Unlike ViT with global self-attention \cite{ViT}, Swin transformer introduces non-overlapping windows and a shifted windowing scheme, enabling cross-window interaction at reduced cost while preserving global context.

Input images are embedded into $d_v$-dimensional patch tokens and passed through hierarchical Swin transformer blocks with patch merging to yield a final feature vector $g_v$. This shared representation is adapted for $\Phi_B$, $\Phi_S$, $\Phi_V$, and $\Phi_M$ via task-specific feedforward networks ${\mathscr{F}}^v_1$, ${\mathscr{F}}^v_s$, ${\mathscr{F}}^v_v$, and ${\mathscr{F}}^v_m$, producing $g^v_1$, $g^v_s$, $g^v_v$, and $g^v_m$, respectively, enabling parameter sharing with task-specific specialization. We first fine-tune the Swin-B variant on a subset of our dataset, then use its weights to initialize end-to-end IMAGINE training, improving convergence and downstream performance.

\subsubsection{Textual Feature Extractor from Blurb}
For textual features, we use the XLNet-Base transformer \cite{XLNet},
chosen for its permutation-based objective, relative positional encoding, and segment recurrence, which together provide richer bidirectional context and long-sequence modeling. XLNet is first fine-tuned on a domain-specific corpus, then trained end-to-end within IMAGINE. Each blurb is encoded into a $d_t$-dimensional vector $g_t$, which is transformed by ${\mathscr{F}}^t_1$, ${\mathscr{F}}^t_s$, ${\mathscr{F}}^t_m$, and ${\mathscr{F}}^t_t$ into task-specific representations $g^t_1$, $g^t_s$, $g^t_m$, and $g^t_t$ for $\Phi_B$, $\Phi_S$, $\Phi_M$, and $\Phi_T$, respectively.

\subsubsection{Textual Feature Extractor from Cover Text}
We also use the fine-tuned XLNet-Base
\cite{XLNet} to extract textual features from the cover text of the book, obtained by an OCR \cite{gemini} from the cover page image.
Similar to the blurb feature extractor, the cover text is encoded into a $d_c$-dimensional representation $g_c$. This representation is passed through a dedicated feedforward network ${\mathscr{F}}^c$, yielding $g^c = {\mathscr{F}}^c(g_c)$, which is used exclusively by the Level-1 classifier $\Phi_B$. Given the limited and often noisy nature of the cover text in most cases, we avoid incorporating this modality into the Level-2 modules or the SGM.

\subsubsection{Feature Extractor from Metadata}
We design a metadata feature extractor ${\mathscr{F}}^m$, by constructing a knowledge graph from training metadata, where nodes represent field values of four entity types: authors, publishers, Level-1 genres (coarse-grained), and Level-2 genres (fine-grained). The graph encodes six types of directed relations based on co-occurrence patterns: (author, publisher), (author, Level-1 genre), (author, Level-2 genre), (publisher, Level-1 genre), (publisher, Level-2 genre), and (Level-1 genre, Level-2 genre). These semantically meaningful edges capture structural relationships, enabling effective multi-relational representation learning.

To embed these heterogeneous entities, we use TransD~\cite{transD}, a translation-based model that projects entities and relations into relation-specific spaces. TransD improves over prior models like TransE and TransR by dynamically modeling entity-relation interactions with fewer parameters~\cite{transD}, reducing overfitting in sparse graphs. 
Once trained, TransD generates entity embeddings that serve as metadata features. For a sample \( M_i \), we extract and aggregate (e.g., via 
average pooling) the embeddings of its associated authors and publishers to form the metadata vector \( g^m \in \mathbb{R}^{d_m} \). If no relevant entities are seen during training, \( g^m \) defaults to a zero vector. 
The vector \( g^m \) is used only in the Level-1 classifier \( \Phi_B \), and is excluded from later modules (e.g., Level-2 refinement, SGM) due to sparse metadata coverage. We train ${\mathscr{F}}^m$ using a margin-based ranking loss~\cite{transD} over observed and synthetic triplets to ensure robust, discriminative representations.

\subsection{Training Strategy for IMAGINE} 
IMAGINE is trained in two sequential phases to ensure both reliable specializations of modality-specific classifiers and effective learning of the selective gating mechanism.
In the first phase, we construct a reliable subset \({\mathcal{D}}'\) from the original training dataset ${\cal{D}}_{train}$. From \({\mathcal{D}}'\), we create three filtered subsets: 
\({\mathcal{D}}_1\) that excludes visually challenged samples and is used to train the visual classifier \(\Phi_V\); 
\({\mathcal{D}}_2\) that excludes textually challenged samples and is used to train the textual classifier \(\Phi_T\), and 
\({\mathcal{D}}_3\) that excludes all samples that are either visually or textually challenged and is used to train the multi-modal classifier \(\Phi_M\). 
Initially, \(\Phi_B\) is trained using \({\mathcal{D}}'\), while \(\Phi_V\), \(\Phi_T\), and \(\Phi_M\) are subsequently trained on \({\mathcal{D}}_1\), \({\mathcal{D}}_2\), and \({\mathcal{D}}_3\), respectively, with \(\Phi_B\) frozen during this phase.
Once these classifiers are trained on their respective high-confidence subsets, we prepare a dataset for training the selective SGM module \(\Phi_S\). Each sample in \({\mathcal{D}}'\) is passed through all three classifiers (\(\psi_M\), \(\psi_V\), and \(\psi_T\)), and the one producing the most accurate prediction is identified. 
A one-hot target vector is then created to represent the best-performing classifier (with dimensions corresponding to \(\psi_M\), \(\psi_V\), and \(\psi_T\)), setting the appropriate index to `1' and the others to `0'. Samples for which none of the classifiers provide a correct prediction are excluded from this training set. \(\Phi_S\) is then trained using this data to learn to select the most reliable classifier per instance.
In the second phase, the complete training dataset ${\cal{D}}_{train}$ is used to jointly train the 
IMAGINE. Initially, \(\Phi_B\), \(\Phi_V\), \(\Phi_T\), and \(\Phi_M\) are updated while collecting examples for the SGM module. 
After a predefined training epoch count, we switch to updating \(\Phi_S\) with collected samples. This process of alternating between updating the SGM module and classifiers allows \(\Phi_S\) to continually adapt to the evolving performance of the classifiers, ensuring robust and dynamic routing of each input to the most suitable modality-specific branch.

\section{Experiments and Discussions}
\label{4sec:result}
This section presents a comprehensive set of experiments to evaluate the performance of IMAGINE. The experiments were conducted using Pytorch 2.1.0
on a system having an Intel(R) Xeon(R) W-1270 processor with 16 CPU cores, 128 GB of RAM, and a 24 GB NVIDIA RTX A5000 GPU. 

\subsection{Database Employed}
\label{subsec:data_emp}
The primary goal of this study is to analyze multi-modal data from books and identify associated multi-label genres. Since no publicly available multi-modal hierarchical book genre dataset exists, to the best of our knowledge, we 
curated a comprehensive dataset containing 11302 book samples, categorized into 6704 fiction and 4598 non-fiction books, each with 1 to 6 genre labels. The dataset includes cover pages, blurbs, metadata (author, publisher), and multi-label genres.

\begin{table}[!t] 
\centering  
\caption{Hierarchical genre-wise count of the dataset} \label{tab1} 
\begin{adjustbox}{width=0.48\textwidth}
    \begin{tabular}{c |l |c |c || c| l |c |c}
    \hline
    \textbf{Class ID} & \textbf{Genre label} & \textbf{Fiction} & \textbf{Non-fiction} & \textbf{Class ID} & \textbf{Genre label} & \textbf{Fiction} & \textbf{Non-fiction} \\ \hline \hline
    1 & Animals \& Wildlife \& Pets & 590 & 235 & 16 & Literature & 2615 & 670 \\ \hline 
    
    \multirow{2}{*}{2} & \multirow{2}{*}{Arts \& Photography} & \multirow{2}{*}{1188} & \multirow{2}{*}{606} & \multirow{2}{*}{17} & Mystery \& Thriller \& & \multirow{2}{*}{2043} & \multirow{2}{*}{404} \\ 
    & & & & & Suspense \& Horror \& Adventure & & \\ \hline 
    
    3 & Business \& Money & 42 & 256 & 18 & Medical & 70 & 165  \\ \hline
    
    4 & {Childrens\textquoteright} Book & 1714 & 298 & 19 & Meta Text & 35 & 88  \\ \hline 
    
    5 & Comics \& Graphic & 364 & 96 & 20 & Mythology \& Religion \& Spirituality & 890 & 748 \\ \hline 
    
    6 & Computers \& Technology & 27 & 92 & 21 & Press \& Media & 138  & 167  \\ \hline 
    
    7 & Cookbooks \& Food \& Wine & 72 & 223 & 22 & Reference \& Language & 97 & 1003  \\ \hline  

    8 & Crafts \& Hobbies \& Home & 40  & 61 & 23 & Romance & 1445 & 80 \\ \hline  
    
    9 & Environment \& Plant & 92 & 337 & 24 & Science \& Math & 419  & 712 \\ \hline

    10 & Family \& Parenting \& Relationships & 208 & 257 & 25 & Self-help \& Motivation & 72 & 630 \\ \hline 

    11 & Fashion \& Lifestyle & 732 & 204 & 26 & Sports \& Outdoors & 183  & 157 \\ \hline 
    
    12 & Health \& Fitness \& Dieting & 32 & 320 & 27 & Teen \& Young Adult & 1712 & 170 \\ \hline

    13 & History & 1677 & 1619 & 28 & Travel & 92 & 393 \\ \hline
    
    14 & Humanities & 537 & 1555 & 29 & Sci-Fi \& Fantasy & 2715 & \--- \\ \hline 
    
    15 & Humor \& Entertainment & 972 & 376 & 30 & Biographies \& Memoir & \--- & 1256 \\  
    \hline
    \end{tabular}
    \end{adjustbox}
    \label{tab:dataset_table}
\end{table}

Table \ref{tab:dataset_table} presents the details of genre labels across both fiction and non-fiction categories, along with their distribution statistics. Most samples are associated with multiple genres, leading to overlapping counts and contributing to a significant class imbalance. For instance, genres like \emph{sci-fi \& fantasy} and \emph{history} are highly frequent, whereas \emph{computer \& technology} and \emph{crafts \& hobbies \& home} are underrepresented. To mitigate this imbalance, we employed data augmentation techniques, which substantially improved the distribution. However, due to the multi-label nature of the dataset and frequent co-occurrence of majority and minority classes within the same samples, a complete balance could not be achieved. 
The details of dataset creation, modality-specific challenges, statistical characteristics, and data preprocessing, including data augmentation, are discussed in Appendix \textcolor{blue}{A}. 


To mitigate the data imbalance issue while ensuring a proportional representation of each genre, the dataset was split into training (${\cal{D}}_{train}$), validation (${\cal{D}}_{val}$), and test (${\cal{D}}_{test}$) sets with a ratio of $8{:}1{:}1$.

\subsection{Evaluation Metrics} 
To evaluate model performance on ${\cal{D}}_{test}$, we report the following metrics for Level-1 (binary) classification: F1-score ($\mathcal{F}$), and accuracy ($\mathcal{A}$), all in percentage.
For Level-2 (multi-label) classification, we report F1-score and balanced accuracy in micro ($\mathcal{F}_{\mu}$, $\mathcal{BA}_{\mu}$), macro (${\mathcal{F}}_m$, ${\mathcal{BA}}_m$), weighted (${\mathcal{F}}_w$, ${\mathcal{BA}}_w$), and sample-based (${\mathcal{F}}_s$, ${\mathcal{BA}}_s$) forms. Additionally, Hamming loss ($\mathcal{HL}$) is used to capture label-wise mismatches.
These different metrics offer complementary perspectives: micro averages emphasize frequent classes, macro treats all classes equally, weighted accounts for class frequency, and sample-based metrics reflect per-instance performance, crucial for multi-label tasks with class imbalance \cite{eval_metrics2}.

\subsection{Comparative Analysis with Baseline}

\input{tables/multimodal_baseline}

\begin{table*}[]
\centering
\caption{Modality and Module ablation study}
\begin{adjustbox}{width=1\textwidth} 
\begin{tabular}{c|c | cc |  cccc ccccc |  cccc ccccc}
\hline
\multirow{2}{*}{\textbf{Modality Study}} & \multirow{2}{*}{\textbf{Model}}  & \multicolumn{2}{c|}{\textbf{Level-1}} &  
 \multicolumn{9}{c|}{\textbf{Level-2: Fiction}} & 
 \multicolumn{9}{c}{\textbf{Level-2: Non-fiction}} \\ 
\cline{3-22}
 & & $\cal{F}\uparrow$ & 
 ${\cal{A}}\uparrow$ & 
 ${\cal{F}}_{\mu}\uparrow$ & 
 ${\cal{BA}}_{\mu}\uparrow$ & 
 ${\cal{F}}_{m}\uparrow$ & 
 ${\cal{BA}}_{m}\uparrow$ &
 ${\cal{F}}_{w}\uparrow$ &
 ${\cal{BA}}_{w}\uparrow$ & 
 ${\cal{F}}_{s}\uparrow$ & 
 ${\cal{BA}}_{s}\uparrow$
 & \multicolumn{1}{c|}{$\cal{HL}\downarrow$} & 
 ${\cal{F}}_{\mu}\uparrow$ & 
 ${\cal{BA}}_{\mu}\uparrow$ & 
 ${\cal{F}}_{m}\uparrow$ & 
 ${\cal{BA}}_{m}\uparrow$ &
 ${\cal{F}}_{w}\uparrow$ &
 ${\cal{BA}}_{w}\uparrow$ & 
 ${\cal{F}}_{s}\uparrow$ & 
 ${\cal{BA}}_{s}\uparrow$ & 
 ${\cal{HL}}\downarrow$ \\ 
 \hline \hline 
Visual& $UM_v$&{87.32} & {86.22} & {68.31} & {79.88}  & {68.08} & {79.46} & {67.80} & {77.95} & {63.43} & {78.96} & {0.0670} & {60.83} & {75.94} & {58.71} & {74.13} & {60.33} & {74.24} & {53.57} & {74.37} & {0.0776} \\ 
\hline


Textual & $UM_d$& {96.43} & {96.06} & {71.98} & {84.15} & {70.41} & {82.42} & {70.53} & {81.46} & {69.91} & {83.97} & {0.0632} & {73.63} & {84.09} & {73.48} & {84.41} & {72.12} & {83.30} & {71.19} & {84.09} & {0.0572} \\ 
\hline 


\multirow{5}{*}{\rotatebox{0}{Multi-modal}} & $MM_{vd}$  & 96.89& 96.62& 74.87& 86.26& 75.57& 85.90 & 74.85 & 83.91 & 72.39 & 85.73 & 0.0592 & 77.29& 86.44& 77.98& 85.96 & 77.19 & 85.14 & 72.38 &85.46 & 0.0483\\ 

 & $MM_{vdc}$  &  98.21& 98.04& 76.33& 87.12& 77.26& 87.20 & 76.26 & 85.01 & 73.91 & 86.59 & 0.0557& 76.77& 86.09& 77.63& 85.52 & 76.75 & 84.78 & 71.47 & 85.01 & 0.0494\\ 

 & $MM_{vdm}$ & 96.92& 96.63& 76.01& 86.98& 77.69& 87.32 & 75.98 & 84.63 & 73.72 & 86.45 & 0.0566& 77.15& 86.33& 78.37& 86.28 & 77.10 & 85.14 & 71.94 & 85.19 & 0.0486\\

 & $\psi_M$  & \textbf{98.51} & \textbf{98.36} & \underline{76.65} & \underline{87.47} & \underline{78.15} & \underline{87.87} & \underline{76.62} & \underline{85.22} & \underline{74.27} & \underline{86.88} & \underline{0.0553} & \underline{78.46} & \underline{87.00} & \underline{79.39} & \underline{86.74} & \underline{78.40} & \underline{85.83} & \underline{73.23} & \underline{85.90} & \underline{0.0457}\\ \hline



\multirow{1}{*}{\textbf{Flatten Structure}} & $MM_{F}$ & -  & - & 64.28 & 80.47  & 62.38 & 78.53 & 62.64 & 77.91 & 62.12 & 80.34 & 0.0799 & 72.30 & 82.78 & 69.11 & 81.25 & 70.51 & 81.37 & 71.11 & 83.91 & 0.0629 \\  \hline


\multicolumn{2}{c|}{\textbf{IMAGINE}} & \textbf{98.51} & \textbf{98.36} & \textbf{79.32} & \textbf{88.76} & \textbf{80.52} & \textbf{89.16} & \textbf{79.43} & \textbf{87.23} & \textbf{77.19} & \textbf{88.16} & \textbf{0.0485} & \textbf{83.74} & \textbf{91.06} & \textbf{84.27} & \textbf{90.92} & \textbf{83.71} & \textbf{90.22} & \textbf{80.43} & \textbf{90.53} & \textbf{0.0358}\\ \hline
 
\hline
\end{tabular}
\end{adjustbox}
\label{tab:ablation}
\end{table*}

To the best of our knowledge, 
there is hardly any prior
work addresses hierarchical book genre classification using a comprehensive multi-modal dataset that jointly leverages cover images, OCR-extracted cover texts, blurbs, and metadata. Existing book recommendation or genre prediction methods are either unimodal or flatten genre hierarchies, limiting their real-world applicability. For a fair benchmarking, we flatten our taxonomy into 58 classes (29 fiction $+$ 29 non-fiction) when adapting baselines.

\emph{(i) Past book genre classifiers}:
We engaged architectures of Scofield et al. \cite{port_book}, Ullah et al. \cite{bangla_book}, N.-Flores et al. \cite{spanish_book}, and evaluated them on our dataset to ensure fair comparison. While \cite{port_book}’s blurb-based approach achieved only moderate performance, \cite{bangla_book}’s CNN-BiLSTM with attention showed improvements but struggled with cross-modality reasoning, and \cite{spanish_book}’s RoBERTa-based classifier, though stronger, failed to capture label dependencies and remained sensitive to domain shift. As shown in Fig.~\ref{fig:comparison}(a), IMAGINE consistently outperformed all these past methods,
underscoring the effectiveness of its multi-modal integration, selective gating, and hierarchical taxonomy for robust book genre classification.

\emph{(ii) Closed-source LLMs}: 
Compared against Gemini-2.5 Flash \cite{gemini}, GPT-4.1 Mini \cite{openai2025gpt4.1mini}, and Deepseek-V3 \cite{deepseek}, IMAGINE consistently yielded higher performance (Fig.~\ref{fig:comparison}(b)). This underscores that domain-specific architecture with hierarchical supervision can surpass general-purpose LLMs.

\emph{(iii) Open-source LLMs}:
Against Gemma2-2B \cite{gemma2}, Qwen3-4B \cite{qwen3}, and Mistral-7B \cite{mistral}, IMAGINE again led across metrics (Fig.~\ref{fig:comparison}(c)),
highlighting the benefit of task-specific selective fusion over generic pretrained architectures.

\emph{(iv) Open-source Vision–Language Models (VLMs)}:
We engaged BLIP \cite{BLIP}, CLIP \cite{CLIP}, AltCLIP \cite{AltCLIP}, ALIGN \cite{ALIGN}, and FLAVA \cite{FLAVA} to exploit both cover image and blurb, but were optimized for alignment tasks rather than structured classification. IMAGINE surpassed all these VLMs
(Fig.~\ref{fig:comparison}(d)), 
showing the effectiveness of its hierarchical routing and multi-modal specialization.

\emph{(v) Hierarchical classifiers}:
Methods such as HiAGM \cite{HiAGM}, HiTiN \cite{HiTiN}, and HILL \cite{hill} explicitly leverage hierarchical structures but operate in single-modality textual settings. While they performed better than flat classifiers, they lacked multi-modal adaptivity. IMAGINE outperformed these classifiers
(Fig.~\ref{fig:comparison}(e)), demonstrating its ability to capture deeper label dependencies and semantic relations through multi-modal supervision.

Across all categories, IMAGINE outperformed baselines by combining hierarchical design, adaptive modality gating, and multi-modal fusion, achieving state-of-the-art performance in book genre classification.
\begin{table*}[!hbt]
    \centering
    \caption{Qualitative analysis for modality ablation}
    \begin{adjustbox}{width=0.91\textwidth}
    \begin{tabular}{c|ll}
    \hline 
    
    \rowcolor[HTML]{D9D9D9}
    \multicolumn{3}{c}{\emph{(a)} \textbf{9780394929132:} Great Day for Up; \textbf{Author:} Dr. Seuss; \textbf{Publisher:} Random House Books for Young Readers
    }\\
    \hline 
    & \\ [\dimexpr-\normalbaselineskip+1.5pt]
    
    \multirow{7}{*}{\includegraphics[width=0.1\linewidth]{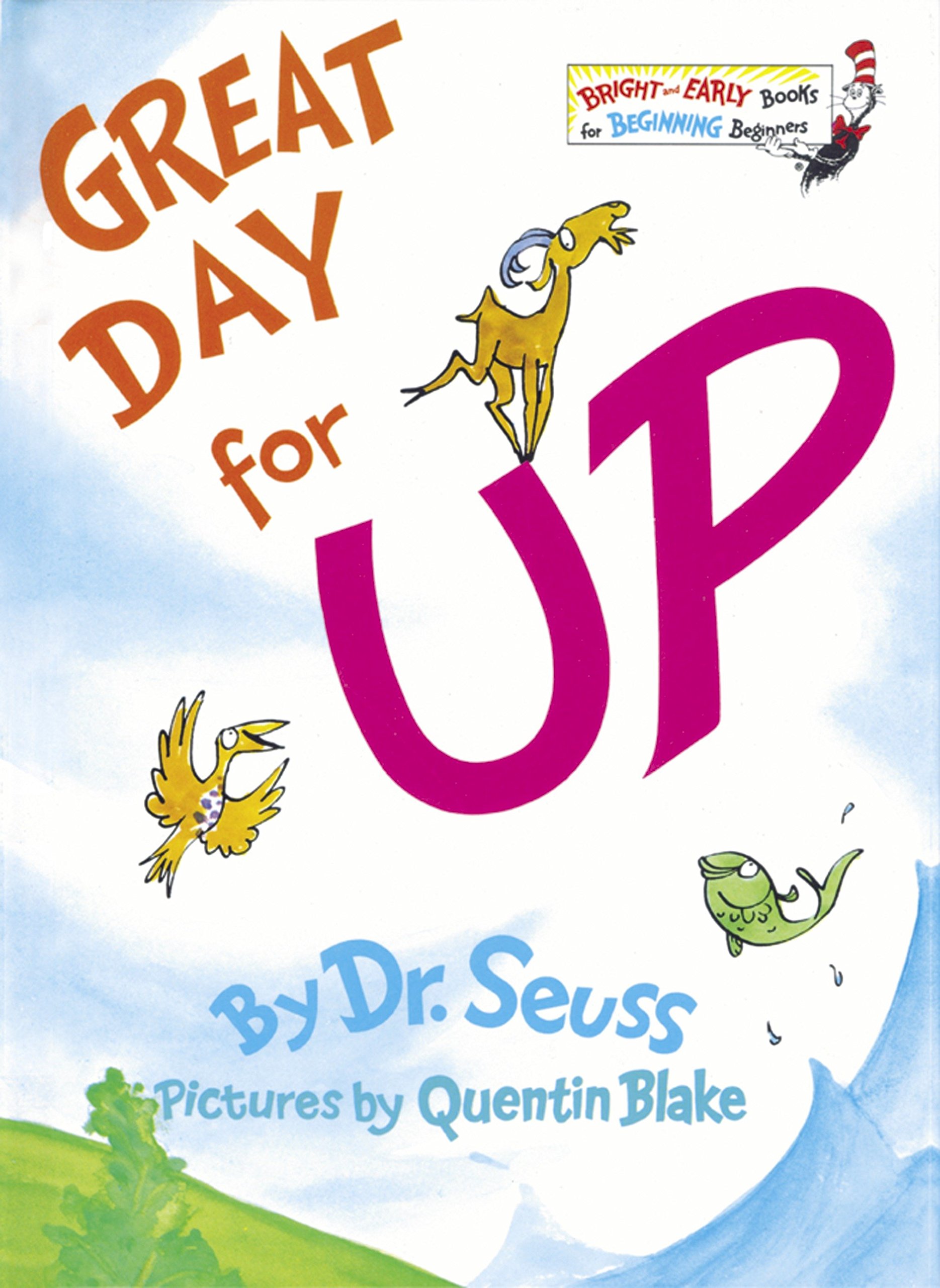}} & \multirow{2}{*}{Blurb} & Up! Up! The sun is getting up. The sun gets up. So UP with you! Discover the different meanings of "up", conveyed with \\
    && {merry verse and illustrations in a happy book that celebrates the joy of life. $\ldots$}\\
    \cline{2-3}
    
    & Actual Genre & (Fiction, \{Arts \& Photography, {Childrens\textquoteright} Book, Humor \& Entertainment, Literature, Sci-Fi \& Fantasy\})\\  \cline{2-3}
    
    & $UM_v$ & (Fiction, \{Arts \& Photography, {Childrens\textquoteright} Book\})\\  
    
    & $UM_d$ & \textcolor{blue}{(Fiction, \{Arts \& Photography, {Childrens\textquoteright} Book, Humor \& Entertainment, Literature, Sci-Fi \& Fantasy\})}$^\ast$\\  
    
    & $MM_{vd}$ & (Fiction, \{Arts \& Photography, {Childrens\textquoteright} Book, Humor \& Entertainment\})\\  
    
    & IMAGINE & \textcolor{blue}{(Fiction, \{Arts \& Photography, {Childrens\textquoteright} Book, Humor \& Entertainment, Literature, Sci-Fi \& Fantasy\})}$^\ast$\\  
    
    \hline \hline
    \rowcolor[HTML]{D9D9D9}
    \multicolumn{3}{c}{\emph{(b)} \textbf{9780553243581:} 
     When Love Dies; \textbf{Author:} Francine Pascal \& Kate William; \textbf{Publisher:} Bantam Books}\\
    \hline 
    & \\ [\dimexpr-\normalbaselineskip+1.5pt]

    \multirow{4}{*}{\includegraphics[width=.1\linewidth]{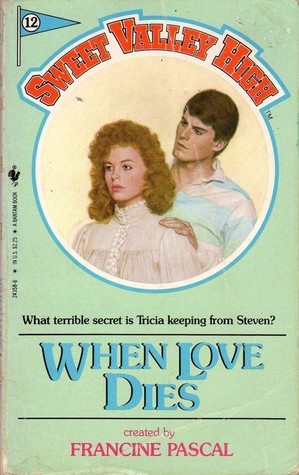}} & \multirow{2}{*}{Blurb} & \multirow{2}{*}{Book by Francine Pascal} \\
    & & \\ \cline{2-3}

    &  Actual Genre & (Fiction, \{Children's Book, Family \& Parenting \& Relationships, Literature, Romance, Teen \& Young Adult\})\\\cline{2-3}

    & $UM_v$ & \textcolor{blue}{(Fiction, \{Children's Book, Family \& Parenting \& Relationships, Literature, Romance, Teen \& Young Adult\})}
    \\ 
    
    & $UM_d$ & (Fiction, \{Arts \& Photography, Humor \& Entertainment, Literature\})
    \\ 
    
    & $MM_{vd}$ & (Fiction, \{Arts \& Photography, Children's Book, Family \& Parenting \& Relationships, Literature, Romance, Teen \& Young Adult\})
    \\  
    
    & IMAGINE & \textcolor{blue}{(Fiction, \{Children's Book, Family \& Parenting \& Relationships, Literature, Romance, Teen \& Young Adult\})}
     \\  
    \hline \hline
    
    \rowcolor[HTML]{D9D9D9}
    \multicolumn{3}{c}{\emph{(c)} \textbf{9780307001504:} 
    The Cat That Climbed the Christmas Tree; \textbf{Author:} Susanne Santoro Whayne \& Christopher Santoro; \textbf{Publisher:} Western Publishing Company Inc}\\
    \hline 
    & \\ [\dimexpr-\normalbaselineskip+1.5pt]

    \multirow{8}{*}{\includegraphics[width=0.1\linewidth]{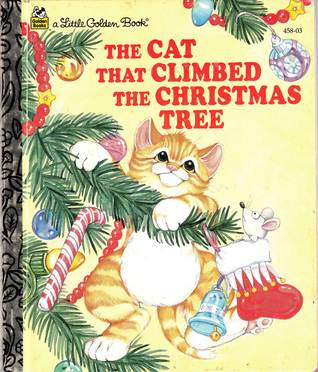}} & \multirow{2}{*}{Blurb} & Benny, the cat, is experiencing his very first Christmas. He eagerly climbs the sparkling Christmas tree. On the way up, \\
    &  & he meets new friends, including a fuzzy reindeer, a velvet mouse, a musical bird and, of course, the lovely angel at the \\
    && top. But how will Benny make it back down the tree? \\ \cline{2-3}

    & Actual Genre & (Fiction, \{ Animals \& Wildlife \& Pets, Arts \& Photography, {Childrens\textquoteright} Book, Mythology \& Religion \& Spirituality\})\\\cline{2-3}
    
    & $UM_v$ & (Fiction, \{Animals \& Wildlife \& Pets, Arts \& Photography, {Childrens\textquoteright} Book\})
    \\ 
    
    & $UM_d$ & (Fiction, \{Animals \& Wildlife \& Pets, {Childrens\textquoteright} Book, Mythology \& Religion \& Spirituality\})
    \\ 
    
    & $MM_{vd}$ & \textcolor{blue}{(Fiction, \{Animals \& Wildlife \& Pets, Arts \& Photography, {Childrens\textquoteright} Book, Mythology \& Religion \& Spirituality\})}
    \\ 
    
    & IMAGINE & \textcolor{blue}{(Fiction, \{Animals \& Wildlife \& Pets, Arts \& Photography, {Childrens\textquoteright} Book, Mythology \& Religion \& Spirituality\})}
    \\  
    
    \hline \hline
    
    \rowcolor[HTML]{D9D9D9}
    \multicolumn{3}{c}{\emph{(d)} \textbf{9781568655802:} Bimbos \& Zombies: Bimbos of the Death Sun / Zombies of the Gene Pool;
    \textbf{Author:} Sharyn McCrumb; \textbf{Publisher:} GuildAmerica Books
    }\\
    \hline 
    & \\ [\dimexpr-\normalbaselineskip+1.5pt]
    
    \multirow{7}{*}{\includegraphics[width=0.1\linewidth]{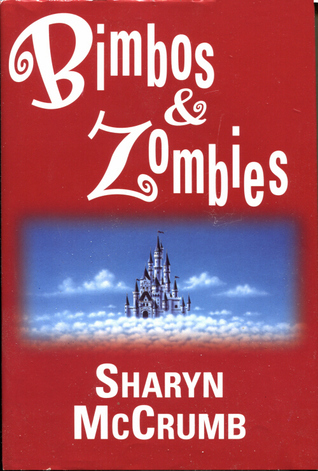}} & \multirow{2}{*}{Blurb} & \multirow{2}{*}{*Bimbos of the Death Sun* Zombies of the Gene Pool }\\ & & \\\cline{2-3}
    
    & Actual Genre & (Fiction, \{Humor \& Entertainment, Mystery \& Thriller \& Suspense \& Horror \& Adventure, Sci-Fi \& Fantasy\})\\  \cline{2-3}
    
    & $UM_v$ & (Fiction, \{History, Mystery \& Thriller \& Suspense \& Horror \& Adventure, Romance\})\\  
    
    & $UM_d$ & (Non-fiction, \{\}) \\  
    
    & $MM_{vd}$ & (Fiction, \{Mystery \& Thriller \& Suspense \& Horror \& Adventure\})\\  
    
    & IMAGINE & \textcolor{blue}{(Fiction, \{Humor \& Entertainment, Mystery \& Thriller \& Suspense \& Horror \& Adventure, Sci-Fi \& Fantasy\})}\\  
    
    \hline \hline
    
    \rowcolor[HTML]{D9D9D9}
    \multicolumn{3}{c}{\emph{(e)} \textbf{9780060186869:} The Blessing of the Animals: True Stories of Ginny, the Dog Who Rescues Cats; \textbf{Author:} Philip Gonz\'alez, \textbf{Publisher:} HarperCollins
    }\\
    \hline 
    & \\ [\dimexpr-\normalbaselineskip+1.5pt]
    
    \multirow{7}{*}{\includegraphics[width=0.1\linewidth]{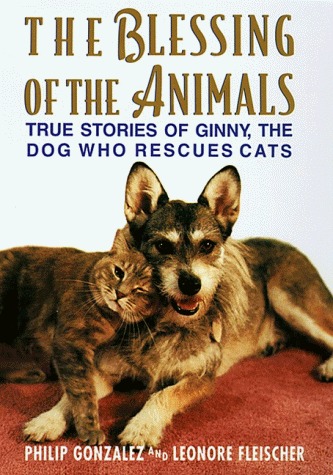}} & \multirow{2}{*}{Blurb} & Many thousands of readers shared the joy of The Dog Who Rescues Cats, the amazing true story of Philip Gonzalez and his \\
    &  & miracle dog, Ginny. Millions more watched their story on television news and talk shows. $\ldots$\\  \cline{2-3}
    
    &Actual Genre & (Non-fiction, \{Animals \& Wildlife \& Pets, Biographies \& Memoir\})\\  \cline{2-3}
    
    &$UM_v$ & (Non-fiction, \{Animals \& Wild life \& Pets, Environment \& Plant, Science \& Math\})\\  
    
    &$UM_d$ & (Non-fiction, \{Animals \& Wildlife \& Pets\})\\ 
    
    &$MM_{vd}$ & (Non-fiction, \{Animals \& Wildlife \& Pets\})\\ 
    
    &IMAGINE & \textcolor{blue}{(Non-fiction, \{Animals \& Wildlife \& Pets, Biographies \& Memoir\})}\\  
    
    \hline
    
    \multicolumn{3}{r}{$^\ast$\textcolor{blue}{Predicted genre set} matches exactly with the actual genre set  }
    \end{tabular}
    \end{adjustbox}

    \label{tab:ablation_qualitative}

\end{table*}


\subsection{Modality Ablation Study} 
In this subsection, we analyze the contribution of each modality for genre prediction by evaluating different ablated versions of IMAGINE:
(a) $UM_v$: an unimodal version of IMAGINE that utilizes only the visual modality extracted from the cover image,
(b) $UM_d$: an unimodal version of IMAGINE that relies solely on the textual modality obtained from the blurb,
(c) $MM_{vd}$: a multi-modal ablated version of IMAGINE that incorporates only the cover image and blurb,
(d) $MM_{vdc}$: a multi-modal ablated version of IMAGINE that considers the cover image, blurb, and 
cover text,
(e) $MM_{vdm}$: a multi-modal ablated version of IMAGINE that includes the cover image, blurb, and metadata,
(f) $\psi_M$: a multi-modal ablated version of IMAGINE that integrates the cover image, blurb, 
cover text, and metadata. The key distinction between $\psi_M$ and IMAGINE is that while IMAGINE dynamically selects between unimodal (visual or textual) and fully multi-modal models, $\psi_M$ strictly relies on multi-modal data without such selective adaptation.

Table \ref{tab:ablation} presents the performance of IMAGINE alongside its various ablated versions, highlighting the importance of incorporating multiple modalities and selectively utilizing them based on their informativeness. From Table \ref{tab:ablation}, we can observe the followings:

\emph{(i) Impact of Unimodal vs. Multi-modal Representations ($UM_v$ / $UM_d$ vs. $MM_{vd}$)}:  Notably, $UM_d$ consistently outperformed $UM_v$ across all evaluation metrics in both level-1 and level-2 classification, indicating that the blurb text provides more informative features for book genre identification than the cover image. Furthermore, $MM_{vd}$ demonstrated a significant performance improvement over both $UM_v$ and $UM_d$ across all evaluation metrics, highlighting that multi-modal features encapsulate richer genre-related information than unimodal features derived solely from the blurb or cover image.

\emph{(ii) Impact of Multi-modalities in IMAGINE}: It is worth noting that $MM_{vdc}$ and $MM_{vdm}$ outperformed $MM_{vd}$ in the majority of cases, indicating that incorporating either metadata or 
cover text enhances feature representation and improves genre identification. Furthermore, $\psi_M$ achieved superior results compared to both $MM_{vdc}$ and $MM_{vdm}$ across all metrics, demonstrating the individual contributions of the cover image, blurb, cover text, 
and metadata. This improvement stems from integrating multiple modalities, leading to richer and more informative feature extraction.

\emph{(iii) Impact of Hierarchical over Flattened Architecture}: We compared $\psi_M$, a hierarchical multi-modal architecture, with $MM_F$, a flattened variant using all modalities but treating the taxonomy as 58 independent classes (29 fiction $+$ 29 non-fiction). $\psi_M$ consistently outperformed $MM_F$, demonstrating that explicitly modeling the hierarchical label dependencies improves classification.

\emph{(iv) Comparison of IMAGINE with its Ablated Versions}: IMAGINE exhibited superior performance, surpassing all
its ablated versions across all evaluation metrics. A key reason for this is the selection mechanism of IMAGINE in level-2 classification, which enables it to dynamically choose the most effective classifier based on the informativeness of the available modalities (cover image and blurb).


Table \ref{tab:ablation_qualitative} highlights the importance of different modalities used by IMAGINE and the significance of its adaptive model selection based on informativeness across modalities.
The table presents multiple examples, including the cover image, blurb (here truncated with “$\ldots$”, when lengthy), and metadata (at the top), along with their actual genre labels and predictions made by $UM_v$, $UM_d$, $MM_{vd}$, and IMAGINE.

In Table \ref{tab:ablation_qualitative}:\emph{(a)}, the cover image conveys limited information, while the blurb is highly informative. Similarly, in Table \ref{tab:ablation_qualitative}:\emph{(b)}, the cover image is more relevant, but the blurb lacks useful details. $MM_{vd}$ struggled to predict all genres accurately, whereas IMAGINE correctly identified them, likely due to its selective adaptation between unimodal and multi-modal models.
In Table \ref{tab:ablation_qualitative}:\emph{(c)}, both the cover image and blurb provide useful information for identifying at least some genres. Here, $MM_{vd}$ correctly predicted the genres, and IMAGINE performed equally well. Notably, IMAGINE consistently outperformed $UM_v$, $UM_d$, and $MM_{vd}$ even when one modality (either the cover image or blurb) lacked sufficient information. This demonstrates IMAGINE's robustness; its ability to effectively leverage complementary features across modalities, and its adaptive model selection based on the informativeness of available modalities.
The importance of cover text and metadata is further emphasized in Table \ref{tab:ablation_qualitative}:\emph{(d)}-\emph{(e)}. In these cases, only IMAGINE successfully predicted the genres, showcasing its superior capability to integrate features extracted from diverse and reliable digital content for accurate genre classification.


\begin{figure}[!t]
    \centering
    \begin{minipage}{0.24\textwidth}
    \centering
    \begin{adjustbox}{width=\textwidth}
    \tiny{(a)} \begin{tikzpicture}[scale=0.60]
            \begin{axis}[
                width=1.32\textwidth,
                height=0.8\textwidth,
                xtick={0, 1, 2, 3},
                xticklabels={$\oplus$, $\odot$, $\otimes$, \rotatebox{90}{$\ominus$}},
                ymin=95, ymax=99,
                ytick={95, 96, 97, 98, 99},
                legend columns=2,
                ymajorgrids=true,
                grid style=dashed,
                colormap/cool,
                every axis label/.append style={font=\tiny},
                xticklabel style={rotate=0, anchor=north, font=\tiny},
                every axis tick label/.append style={font=\tiny},
                legend style={font=\tiny, draw=none, fill=none, at={(0.9, 0.25)}}
            ]
            \addplot[color=blue, mark=triangle] coordinates {
            (0,96.35) (1,96.48) (2,96.67) (3,98.51)};
            \addlegendentry{${\mathcal{F}}$}
            
            \addplot[color=orange, mark=*] coordinates {
            (0,95.98) (1,96.16) (2,96.26) (3,98.36)};
            \addlegendentry{${\mathcal{A}}$}
            \end{axis}
        \end{tikzpicture}
    \end{adjustbox}
     \label{fig:level1_feature_fusion}
    \end{minipage}
    \hfill
      \begin{minipage}{0.24\textwidth}
        \centering
     \begin{adjustbox}{width=\textwidth}
    \small{(b)} \begin{tikzpicture}
            \begin{axis}[
                  width=1.35\textwidth,
                  height=0.8\textwidth,
                  ymin=0, ymax=100,
                  symbolic x coords={A,B,C,D,E,F,IMAGINE},
                  xtick=data,
                  xticklabels={{$(\oplus{,}\otimes)$},{$(\oplus{,}\odot)$},{$(\oplus{,}\ominus)$},
                               {$(\ominus{,}\otimes)$},{$(\ominus{,}\odot)$},{$(\ominus{,}\ominus)$},{\tiny{IMAGINE}}},
                  x tick label style={rotate=90, anchor=east},
                  legend style={font=\tiny, draw=none, fill=none, at={(0.6,0.35)},anchor=north,legend columns=2},
                  enlarge x limits=0.01,
                  every axis label/.append style={font=\tiny},
                  every axis tick label/.append style={font=\tiny},                ]
                \addplot[color=blue, mark=triangle] coordinates {(A,38.97) (B,51.31) (C,58.38) (D,33.69) (E,43.48) (F,67.02) (IMAGINE,80.52)};
                \addlegendentry{F: ${\mathcal{F}_m}$}
                \addplot[color=orange, mark=*] coordinates {(A,66.19) (B,71.93) (C,74.88) (D,62.85) (E,67.77) (F,80.73) (IMAGINE,89.16)};
                \addlegendentry{F: ${\mathcal{BA}_m}$}
                \addplot[color=darkred, mark=triangle] coordinates {(A,45.87) (B,26.36) (C,65.78) (D,42.68) (E,42.56) (F,73.25) (IMAGINE,84.27)};
                \addlegendentry{NF: ${\mathcal{F}_m}$}
                \addplot[color=violet, mark=*] coordinates {(A,71.23) (B,59.98) (C,80.44) (D,69.68) (E,67.37) (F,83.97) (IMAGINE,90.92)};
                \addlegendentry{NF: ${\mathcal{BA}_m}$}
            \end{axis}
        \end{tikzpicture}
        \end{adjustbox}
        \label{fig:level2_feature_fusion}
    \end{minipage}
\caption{Impact of fusions $f_1$, $f_2$, and $f_3$: 
(a) Level-1 ($f_1$), 
(b) Level-2 ($f_2$, $f_3$).
$\oplus$ : addition, 
$\odot$ : self-attention, 
$\otimes$ : cross-attention,
\rotatebox{90}{$\ominus$} : concatenation. 
F: Fiction, NF: Non-fiction. 
}
    \label{fig:feature_fusion}
\end{figure}

\subsection{Impact of Various Fusion Strategies}
Fig. \ref{fig:feature_fusion} examines the impact of different fusion strategies on genre identification performance. 
This figure presents the performance of $\psi_M$ engaging the fusion strategies used in $f_1$, $f_2$, and $f_3$.  
Fig.s \ref{fig:feature_fusion}(a), (b) show that concatenation (\rotatebox{90}{$\ominus$}) consistently achieved the highest performance in both Level-1 and Level-2 classification, outperforming addition ($\oplus$), self-attention ($\odot$), and cross-attention ($\otimes$). Notably, self-attention and cross-attention led to a significant drop in model performance, indicating their inefficacy in this context.

\subsection{Genre-wise Analysis}
\label{App_C:GenreAnalysis}


\begin{figure}[!b]
\begin{subfigure}{0.24\textwidth}
    \centering
\begin{tikzpicture}[scale=0.35]

\foreach \r/\label in {1.25/55,2.5/70,3.75/85,5/100} {
  \draw[gray!20] (0,0) circle (\r);
  \node[gray!50, font=\tiny] at (\r-0.2,0) {\label};
}
\node[gray!50, font=\tiny] at (0.2,0.2) {40};

\foreach \i in {1,...,29} {
  \draw[gray!40] (0,0) -- ({360/29 * (\i-1)}:5);
  \node[font=\tiny, anchor=center] at ({360/29*(\i-1)}:5.4) {\i};
}

\newcommand{\radarpoint}[2]{({360/29*(#1-1)}:{(#2-40)/12})}

\draw[black]
  \radarpoint{1}{71.62} -- \radarpoint{2}{70.00} -- \radarpoint{3}{84.48} --
  \radarpoint{4}{86.19} -- \radarpoint{5}{74.19} -- \radarpoint{6}{80.77} --
  \radarpoint{7}{90.12} -- \radarpoint{8}{94.64} -- \radarpoint{9}{82.88} --
  \radarpoint{10}{67.65} -- \radarpoint{11}{60.00} -- \radarpoint{12}{93.24} --
  \radarpoint{13}{70.53} -- \radarpoint{14}{72.55} -- \radarpoint{15}{74.18} --
  \radarpoint{16}{80.43} -- \radarpoint{17}{89.08} -- \radarpoint{18}{80.53} --
  \radarpoint{19}{91.86} -- \radarpoint{20}{71.56} -- \radarpoint{21}{85.29} --
  \radarpoint{22}{94.92} -- \radarpoint{23}{68.14} -- \radarpoint{24}{68.92} --
  \radarpoint{25}{94.51} -- \radarpoint{26}{82.14} -- \radarpoint{27}{69.54} --
  \radarpoint{28}{86.96} -- \radarpoint{29}{84.54} -- cycle;

\draw[darkgreen]
  \radarpoint{1}{78.52} -- \radarpoint{2}{80.00} -- \radarpoint{3}{85.96} --
  \radarpoint{4}{86.67} -- \radarpoint{5}{64.79} -- \radarpoint{6}{98.44} --
  \radarpoint{7}{89.02} -- \radarpoint{8}{94.64} -- \radarpoint{9}{90.20} --
  \radarpoint{10}{86.25} -- \radarpoint{11}{76.03} -- \radarpoint{12}{94.52} --
  \radarpoint{13}{82.11} -- \radarpoint{14}{68.94} -- \radarpoint{15}{77.07} --
  \radarpoint{16}{83.38} -- \radarpoint{17}{69.58} -- \radarpoint{18}{91.92} --
  \radarpoint{19}{87.78} -- \radarpoint{20}{76.26} -- \radarpoint{21}{65.91} --
  \radarpoint{22}{82.35} -- \radarpoint{23}{80.08} -- \radarpoint{24}{69.86} --
  \radarpoint{25}{90.53} -- \radarpoint{26}{80.23} -- \radarpoint{27}{76.92} --
  \radarpoint{28}{77.92} -- \radarpoint{29}{80.55} -- cycle;

\draw[darkred]
  \radarpoint{1}{74.91} -- \radarpoint{2}{74.67} -- \radarpoint{3}{85.22} --
  \radarpoint{4}{86.43} -- \radarpoint{5}{69.17} -- \radarpoint{6}{88.73} --
  \radarpoint{7}{89.57} -- \radarpoint{8}{94.64} -- \radarpoint{9}{86.38} --
  \radarpoint{10}{75.82} -- \radarpoint{11}{67.07} -- \radarpoint{12}{93.88} --
  \radarpoint{13}{75.88} -- \radarpoint{14}{70.70} -- \radarpoint{15}{75.60} --
  \radarpoint{16}{81.88} -- \radarpoint{17}{78.13} -- \radarpoint{18}{85.85} --
  \radarpoint{19}{89.77} -- \radarpoint{20}{73.84} -- \radarpoint{21}{74.36} --
  \radarpoint{22}{88.19} -- \radarpoint{23}{73.63} -- \radarpoint{24}{69.39} --
  \radarpoint{25}{92.47} -- \radarpoint{26}{81.18} -- \radarpoint{27}{73.04} --
  \radarpoint{28}{82.19} -- \radarpoint{29}{82.50} -- cycle;

\draw[blue]
  \radarpoint{1}{87.79} -- \radarpoint{2}{86.27} -- \radarpoint{3}{92.68} --
  \radarpoint{4}{91.26} -- \radarpoint{5}{81.86} -- \radarpoint{6}{98.72} --
  \radarpoint{7}{94.24} -- \radarpoint{8}{97.22} -- \radarpoint{9}{94.45} --
  \radarpoint{10}{92.02} -- \radarpoint{11}{85.41} -- \radarpoint{12}{97.09} --
  \radarpoint{13}{86.28} -- \radarpoint{14}{82.98} -- \radarpoint{15}{86.52} --
  \radarpoint{16}{83.91} -- \radarpoint{17}{83.04} -- \radarpoint{18}{95.21} --
  \radarpoint{19}{93.65} -- \radarpoint{20}{85.94} -- \radarpoint{21}{82.79} --
  \radarpoint{22}{91.08} -- \radarpoint{23}{86.47} -- \radarpoint{24}{84.16} --
  \radarpoint{25}{95.09} -- \radarpoint{26}{89.61} -- \radarpoint{27}{84.91} --
  \radarpoint{28}{88.66} -- \radarpoint{29}{86.29} -- cycle;

\draw[orange]
  \radarpoint{1}{97.07} -- \radarpoint{2}{92.54} -- \radarpoint{3}{99.40} --
  \radarpoint{4}{95.86} -- \radarpoint{5}{98.93} -- \radarpoint{6}{99.00} --
  \radarpoint{7}{99.46} -- \radarpoint{8}{99.80} -- \radarpoint{9}{98.70} --
  \radarpoint{10}{97.78} -- \radarpoint{11}{97.79} -- \radarpoint{12}{99.67} --
  \radarpoint{13}{90.46} -- \radarpoint{14}{97.01} -- \radarpoint{15}{95.96} --
  \radarpoint{16}{84.44} -- \radarpoint{17}{96.49} -- \radarpoint{18}{98.50} --
  \radarpoint{19}{99.53} -- \radarpoint{20}{95.62} -- \radarpoint{21}{99.67} --
  \radarpoint{22}{99.80} -- \radarpoint{23}{92.86} -- \radarpoint{24}{98.46} --
  \radarpoint{25}{99.66} -- \radarpoint{26}{98.99} -- \radarpoint{27}{92.89} --
  \radarpoint{28}{99.40} -- \radarpoint{29}{92.04} -- cycle;

\begin{scope}[shift={(0,-6.5)}] 
  \matrix[
    matrix of nodes,
    nodes={anchor=west, font=\tiny},
    column sep=0pt, row sep=0pt
  ]{
    \draw[black, thick] (0,0)--(0.2,0); & $\mathcal{P}$   & 
    \draw[darkgreen, thick] (0,0)--(0.2,0); & $\mathcal{R}$ &
    \draw[darkred, thick] (0,0)--(0.2,0); & $\mathcal{F}$ &
    \draw[blue, thick] (0,0)--(0.2,0); & $\mathcal{BA}$  &
    \draw[orange, thick] (0,0)--(0.1,0); & $\mathcal{S}_p$ & & \\
  };
\end{scope}

\end{tikzpicture}
\caption{\emph{Fiction}}
\end{subfigure}
\hfill
\begin{subfigure}{0.24\textwidth}
    \centering
\begin{tikzpicture}[scale=0.35]

\foreach \r/\label in {1.25/55,2.5/70,3.75/85,5/100} {
  \draw[gray!20] (0,0) circle (\r);
  \node[gray!50, font=\tiny] at (\r-0.2,0) {\label};
}
\node[gray!50, font=\tiny] at (0.2,0.2) {40};

\foreach \i in {1,...,28} {
  \draw[gray!40] (0,0) -- ({360/29 * (\i-1)}:5);
  \node[font=\tiny, anchor=center] at ({360/29*(\i-1)}:5.4) {\i};
}
\foreach \i in {29} {
  \draw[gray!40] (0,0) -- ({360/29 * (\i-1)}:5);
  \node[font=\tiny, anchor=center] at ({360/29*(\i-1)}:5.4) {30};
}

\newcommand{\radarpoint}[2]{({360/29*(#1-1)}:{(#2-40)/12})}

\draw[black]
  \radarpoint{1}{85.71} -- \radarpoint{2}{78.16} -- \radarpoint{3}{84.06} --
  \radarpoint{4}{87.21} -- \radarpoint{5}{96.59} -- \radarpoint{6}{92.86} --
  \radarpoint{7}{91.07} -- \radarpoint{8}{81.82} -- \radarpoint{9}{76.70} --
  \radarpoint{10}{94.79} -- \radarpoint{11}{86.41} -- \radarpoint{12}{90.65} --
  \radarpoint{13}{84.46} -- \radarpoint{14}{86.61} -- \radarpoint{15}{91.82} --
  \radarpoint{16}{81.25} -- \radarpoint{17}{83.13} -- \radarpoint{18}{82.61} --
  \radarpoint{19}{94.44} -- \radarpoint{20}{75.19} -- \radarpoint{21}{87.34} --
  \radarpoint{22}{80.94} -- \radarpoint{23}{91.67} -- \radarpoint{24}{78.28} --
  \radarpoint{25}{77.60} -- \radarpoint{26}{84.31} -- \radarpoint{27}{84.31} --
  \radarpoint{28}{80.21} -- \radarpoint{29}{73.84} -- cycle;

\draw[darkgreen]
  \radarpoint{1}{87.80} -- \radarpoint{2}{84.47} -- \radarpoint{3}{79.45} --
  \radarpoint{4}{78.95} -- \radarpoint{5}{84.16} -- \radarpoint{6}{92.86} --
  \radarpoint{7}{91.07} -- \radarpoint{8}{88.24} -- \radarpoint{9}{92.94} --
  \radarpoint{10}{90.10} -- \radarpoint{11}{84.76} -- \radarpoint{12}{88.18} --
  \radarpoint{13}{85.12} -- \radarpoint{14}{82.91} -- \radarpoint{15}{80.16} --
  \radarpoint{16}{76.92} -- \radarpoint{17}{75.00} -- \radarpoint{18}{78.35} --
  \radarpoint{19}{90.43} -- \radarpoint{20}{80.00} -- \radarpoint{21}{81.18} --
  \radarpoint{22}{89.31} -- \radarpoint{23}{95.65} -- \radarpoint{24}{85.64} --
  \radarpoint{25}{85.14} -- \radarpoint{26}{89.58} -- \radarpoint{27}{52.44} --
  \radarpoint{28}{81.91} -- \radarpoint{29}{86.31} -- cycle;

\draw[darkred]
  \radarpoint{1}{86.75} -- \radarpoint{2}{81.19} -- \radarpoint{3}{81.69} --
  \radarpoint{4}{82.87} -- \radarpoint{5}{89.95} -- \radarpoint{6}{92.86} --
  \radarpoint{7}{91.07} -- \radarpoint{8}{84.91} -- \radarpoint{9}{84.04} --
  \radarpoint{10}{92.39} -- \radarpoint{11}{85.58} -- \radarpoint{12}{89.40} --
  \radarpoint{13}{84.79} -- \radarpoint{14}{84.72} -- \radarpoint{15}{85.59} --
  \radarpoint{16}{79.03} -- \radarpoint{17}{78.86} -- \radarpoint{18}{80.42} --
  \radarpoint{19}{92.39} -- \radarpoint{20}{77.52} -- \radarpoint{21}{84.15} --
  \radarpoint{22}{84.92} -- \radarpoint{23}{93.62} -- \radarpoint{24}{81.80} --
  \radarpoint{25}{81.20} -- \radarpoint{26}{86.87} -- \radarpoint{27}{64.66} --
  \radarpoint{28}{81.05} -- \radarpoint{29}{79.59} -- cycle;

\draw[blue]
  \radarpoint{1}{93.39} -- \radarpoint{2}{90.48} -- \radarpoint{3}{89.26} --
  \radarpoint{4}{88.99} -- \radarpoint{5}{91.95} -- \radarpoint{6}{96.26} --
  \radarpoint{7}{95.32} -- \radarpoint{8}{93.70} -- \radarpoint{9}{95.43} --
  \radarpoint{10}{94.83} -- \radarpoint{11}{91.77} -- \radarpoint{12}{93.65} --
  \radarpoint{13}{89.07} -- \radarpoint{14}{88.44} -- \radarpoint{15}{89.68} --
  \radarpoint{16}{87.06} -- \radarpoint{17}{86.89} -- \radarpoint{18}{88.48} --
  \radarpoint{19}{94.99} -- \radarpoint{20}{88.52} -- \radarpoint{21}{90.16} --
  \radarpoint{22}{91.45} -- \radarpoint{23}{97.48} -- \radarpoint{24}{90.51} --
  \radarpoint{25}{90.56} -- \radarpoint{26}{94.46} -- \radarpoint{27}{75.87} --
  \radarpoint{28}{90.13} -- \radarpoint{29}{87.98} -- cycle;

\draw[orange]
  \radarpoint{1}{98.97} -- \radarpoint{2}{96.48} -- \radarpoint{3}{99.06} --
  \radarpoint{4}{99.04} -- \radarpoint{5}{99.74} -- \radarpoint{6}{99.66} --
  \radarpoint{7}{99.58} -- \radarpoint{8}{99.16} -- \radarpoint{9}{97.93} --
  \radarpoint{10}{99.56} -- \radarpoint{11}{98.77} -- \radarpoint{12}{99.12} --
  \radarpoint{13}{93.02} -- \radarpoint{14}{93.96} -- \radarpoint{15}{99.19} --
  \radarpoint{16}{97.20} -- \radarpoint{17}{98.78} -- \radarpoint{18}{98.60} --
  \radarpoint{19}{99.56} -- \radarpoint{20}{97.05} -- \radarpoint{21}{99.14} --
  \radarpoint{22}{93.59} -- \radarpoint{23}{99.30} -- \radarpoint{24}{95.38} --
  \radarpoint{25}{95.97} -- \radarpoint{26}{99.33} -- \radarpoint{27}{99.31} --
  \radarpoint{28}{98.34} -- \radarpoint{29}{89.66} -- cycle;

\begin{scope}[shift={(0,-6.5)}] 
  \matrix[
    matrix of nodes,
    nodes={anchor=west, font=\tiny},
    column sep=0pt, row sep=0pt
  ]{
    \draw[black, thick] (0,0)--(0.2,0); & $\mathcal{P}$   & 
    \draw[darkgreen, thick] (0,0)--(0.2,0); & $\mathcal{R}$ &
    \draw[darkred, thick] (0,0)--(0.2,0); & $\mathcal{F}$ &
    \draw[blue, thick] (0,0)--(0.2,0); & $\mathcal{BA}$  &
    \draw[orange, thick] (0,0)--(0.1,0); & $\mathcal{S}_p$ & & \\
  };
\end{scope}

\end{tikzpicture}
\caption{\emph{Non-fiction} }
\end{subfigure}
\caption{Genre-wise performance analysis}
\label{fig:genrewise_analysis}
\end{figure}

We present genre-wise performances across 
precision ($\cal{P}$),
recall ($\cal{R}$),
F1-score ($\cal{F}$), 
balanced accuracy ($\cal{BA}$), and 
specificity (${\cal{S}}p$) in Fig.~\ref{fig:genrewise_analysis}, with the following analysis focusing on $\cal{F}$. 
%
For fiction (ID: 1--29), the mean $\mathcal{F}$ = 80.52 (median = 81.18, standard deviation = 7.87) indicates higher variance, suggesting larger inter-genre fluctuations than in non-fiction. Top-performing fiction genres include \emph{crafts \& hobbies \& home} (ID:~8, $\mathcal{F}$ = 94.64), \emph{health \& fitness \& dieting} (ID:~12, $\mathcal{F}$ = 93.88), \emph{self-help \& motivation} (ID:~25, $\mathcal{F}$ = 92.47), \emph{meta text} (ID:~19, $\mathcal{F}$ = 89.77), and \emph{cookbooks \& food \& wine} (ID:~7, $\mathcal{F}$ = 89.57), reflecting strong alignment of multi-modal cues (cover image + blurbs) with hierarchical supervision. In contrast, weaker fiction genres such as \emph{fashion \& lifestyle} (ID:~11, $\mathcal{F}$ = 67.07), \emph{comics \& graphic} (ID:~5, $\mathcal{F}$ = 69.17), \emph{science \& math} (ID:~24, $\mathcal{F}$ = 69.39), \emph{humanities} (ID:~14, $\mathcal{F}$ = 70.70), and \emph{teen \& young adult} (ID:~27, $\mathcal{F}$ = 73.04) suffer from data sparsity and higher semantic confusability at leaf levels.

For non-fiction (IDs 1--28, 30), the mean $\mathcal{F}=84.27$ (median = 84.72, standard deviation = 5.82) shows stronger uniformity. Leading genres such as \emph{romance} (ID:~23, $\mathcal{F}$ = 93.62), \emph{computers \& technology} (ID:~6, $\mathcal{F}$ = 92.86), \emph{meta text} (ID:~19, $\mathcal{F}$ = 92.39), \emph{family \& parenting \& relationships} (ID:~10, $\mathcal{F}$ = 92.39), and \emph{cookbooks \& food \& wine} (ID:~7, $\mathcal{F}$ = 91.07) benefit from abundant textual-visual alignment. Meanwhile, low-resource or niche genres such as \emph{teen \& young adult} (ID:~27, $\mathcal{F}$ = 64.66), \emph{mythology \& religion \& spirituality} (ID:~20, $\mathcal{F}$ = 77.52), \emph{mystery \& thriller \& suspense \& horror \& adventure} (ID:~17, $\mathcal{F}$ = 78.86), \emph{literature} (ID:~16, $\mathcal{F}$ = 79.03), and \emph{biographies \& memoir} (ID:~30, $\mathcal{F}$ = 79.59) exhibit lower scores due to limited samples. Overall, IMAGINE achieves high and stable $\mathcal{F}$ in majority of the genres, particularly in non-fiction, by exploiting hierarchical supervision and multi-modal gating, while performance dips in minor, underrepresented genres highlight the need for augmentation or reweighting strategies. This asymmetry between fiction (larger peaks but deeper troughs) and non-fiction (more stable gains) reflects the combined influence of label frequency and semantic ambiguity as noted in Table~\ref{tab:dataset_table}.

Appendix \textcolor{blue}{B} presents framework analysis with different feature extractor, while Appendix \textcolor{blue}{C} provides additional qualitative results. Overall, the findings show that IMAGINE consistently outperforms multi-modal baselines and unimodal models, highlighting its effectiveness in hierarchical multi-label genre classification using multi-modal inputs.


\section{Conclusion}
\label{5sec:conclusion}
This paper presents IMAGINE, a novel multi-modal framework for hierarchical book genre classification that integrates book covers, blurbs, and metadata for robust and accurate predictions. By leveraging a two-level hierarchy, IMAGINE effectively addresses multi-label dependencies and data imbalance. Experiments show its clear advantage over state-of-the-art methods. This research underscores the value of multi-modal integration and structured genre hierarchies in improving book recommendations and content organization. Future work will focus on expanding to finer sub-genres and supporting deeper hierarchies
with sustained performance. 

\bibliographystyle{IEEEtran}  
\bibliography{ref.bib} 

\section*{Supplementary Appendix}
Appendices A, B, C are provided in \href{https://github.com/Utsav30/IMAGINE}{https://github.com/Utsav30/IMAGINE}.

\end{document}